# Experimental Study of a Vortex Spin-Torque Oscillator in an MTJ with a Vortex Polarizer

*Maksim Stebliy*, *Alex Jenkins, Luana Benetti, Ricardo Ferreira*

*INL - International Iberian Nanotechnology Laboratory, Braga 4715-330, Portugal*

*maksim.steblii@inl.int*

Abstract

*Spin-torque nano-oscillators (STNOs) are promising nanoscale microwave sources for spintronic applications, serving as signal generators or elements in neuromorphic computing systems. In this paper, we investigate the experimental realization of an oscillator based on a magnetic tunnel junction (MTJ) comprising two magnetic layers: a reference layer (RL) and a free layer (FL). We demonstrate that, when magnetic vortices with opposite chirality and polarity are formed in layers, the application of a current induces auto-oscillations even in the absence of external magnetic fields. This effect is observed in devices with diameters ranging from 800 to 1000 nm, exhibiting oscillation frequencies between 110 and 60 MHz. The underlying mechanism is attributed to the action of a spin current with vortex-like polarization, injected from the RL, interacting with the magnetic vortex in the FL. This interaction generates a local out-of-plane effective field due to spin-transfer torque, which acts on the vortex core and initiates its motion. The observed mechanism differs qualitatively from the case of uniformly polarized spin currents perpendicular to the plane, where the resulting in-plane field acts on the planar components of the vortex magnetization.*

1. Introduction

Spin-torque nano-oscillators (STNOs) are a promising class of spintronic devices that exploit spin-transfer torque (STT) to induce and sustain auto-oscillations of magnetic moments [1] [2]. Their nanoscale dimensions, wide frequency tunability (ranging from MHz to sub-THz), and compatibility with CMOS technology make them highly attractive for applications such as on-chip signal generation [3] [4], neuromorphic computing [5], and wireless communication [6]. STNOs offer several advantages, including low power consumption, rich nonlinear dynamics, and the potential for mutual synchronization, positioning them as key components for next-generation information processing systems.

Various types of STNOs have been developed, including those based on spin valves [7], magnetic tunnel junctions (MTJs) [8] [9], and spin-Hall geometries [10]. Depending on the magnetization configuration and excitation mode, STNOs can support diverse dynamic states such as propagating spin waves, bullet modes, magnetic droplets, and vortices [1]. Among these, vortex-based STNOs have attracted particular interest due to their low excitation threshold, high output power, and narrow spectral linewidth [11] [12] [13].

A magnetic vortex oscillator operates by passing an electric current vertically through a structure composed of two magnetic layers. One layer serves as the polarizer, determining the spin polarization of the current, while the second layer hosts a magnetic vortex, whose in-plane magnetisation forms increasingly small closed loops, apart from an out-of-plane magnetic vortex located at the centre – core region. The spin polarised current form the polariser results in sustained oscillations of the vortex in the free layer due to its interaction with the spin current. In most cases, vortex oscillators rely on a polarizer in a single-domain state, with magnetization oriented either perpendicularly either via anisotropy [14] [15] or via the presence of a large perpendicular field [16] [17] - conditions that are often impractical for real-world applications.

An alternative approach involves exciting oscillations via a spin current with a non-uniform polarization distribution, achievable when the polarizer itself contains a magnetic vortex. This phenomenon was previously observed in spin valves with an unpinned reference layer [18] [19] [20] [21]. Recently, we investigated the same mechanism in a three-layer MTJ, where two antiparallel RKKY coupled pinning and reference layers form synthetic antiferromagnet. The reference layer acts as the pinned polarizer and the vortex oscillates in the free layer [arXiv].

In this report, we use a two-layer MTJ, where single layer is directly coupled to an antiferromagnet, which acts as both the pinned and reference layer, capable of generating a non-homogeneous spin current due to the presence of a "pinned vortex". The spin current generated by this pinned vortex results in a steady state auto-oscillation of the "free vortex" in the free layer, however, the existence of these auto-oscillations is strongly affected by the dipolar coupling between the free and pinned vortices as well as any anisotropy which exists due to the exchange coupling with the antiferromagnet.

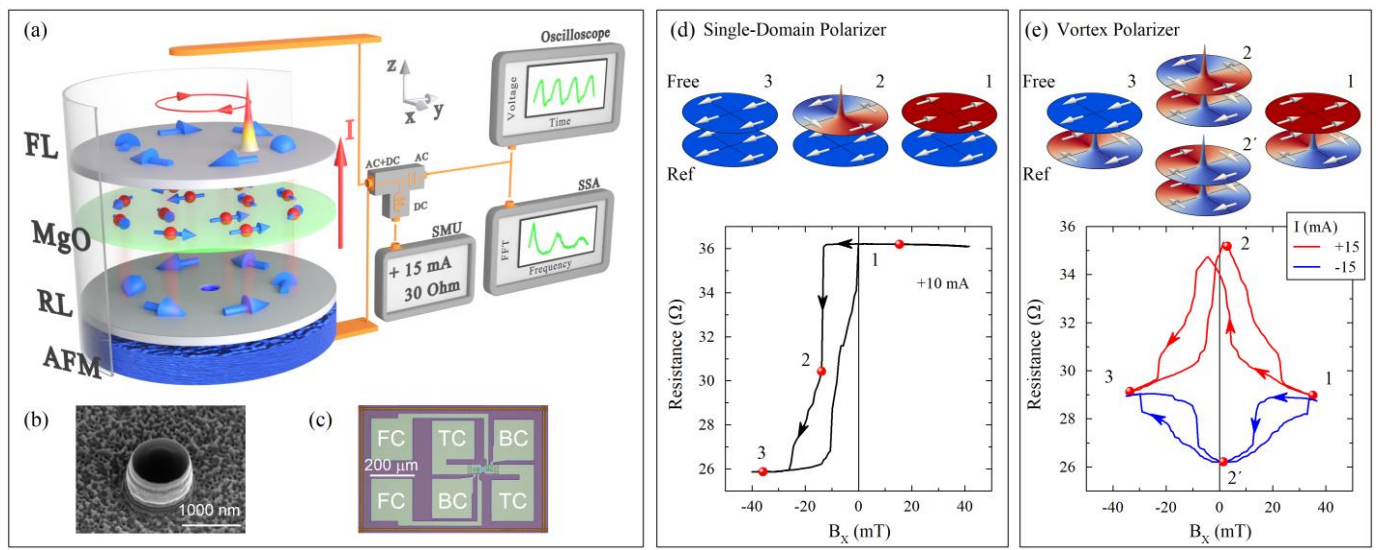

*Fig. 1.* (a) Schematic illustration of the MTJ device with electrical connections under study. Images of the MTJ column at the base of the device, obtained using a scanning electron microscope, and the final device with electrical contacts are shown at the bottom. Hysteresis loops obtained from resistance

*measurements during magnetic field sweeps for cases where a single-domain state (b) or a vortex state (c) is fixed in the reference layer, depending on the parameters of the field-cooling procedure. The main stages of the magnetization reversal process are marked with red dots, with corresponding illustrations of the magnetic structures in the free and reference layers.*

## 2. Description of the device and concept

Experimental studies were conducted on magnetic tunnel junctions with pillar diameters ranging from 600 to 1000 nm, Fif.1b. The layer stack used was IrMn(10)/CoFe(2)/CoFeSiB(16)/CoFeB(2.5)/MgO/CoFeB(2)/Ta(0.21)/NiFe(7), Fig. 1a. Full details of the stack composition and sample preparation are provided in the *Materials and Methods* section, while the magnetic parameters are listed in S.1. Functionally, the structure consists of two magnetic layers - the reference layer (RL) and the free layer (FL), as shown in Fig. 1a - separated by an MgO spacer. This spacer contributes a tunnel magnetoresistance (TMR) of 120%, with a resistance-area product of 9 $\Omega \cdot \mu m^2$.

The magnetic behaviour of the MTJ was investigated under two conditions. In the first case, a field-cooling procedure was performed in the presence of a 1 T in-plane magnetic field to stabilize a single-domain state in the RL. In the second case, field cooling was carried out without an external magnetic field, resulting in vortex formation in the RL due to thermal relaxation.

To verify the magnetic configuration of the devices, resistance measurements were performed using a source-measure unit connected to the top and bottom contacts during a magnetic field sweep. High-frequency processes were investigated through the same contacts via a bias-tee, with signal detection carried out using an oscilloscope and a spectrum analyser (Fig. 1a). The magnetic field along the x-axis was generated by current flowing through a field line placed above the MTJ pillar (Fig. 1c), while the field along the y-axis was produced by an electromagnet.

## 3. Results

The results of the study are presented in the following sequence. In Chapter 3.1, the fixation of a magnetic vortex in the reference layer is demonstrated by comparing hysteresis loops with those corresponding to a single-domain state. Chapter 3.2 investigates the conditions for the excitation of oscillations depending on the mutual chirality and polarity of the vortices. In Chapter 3.3, the influence of the external magnetic field on the time-domain of the oscillations is analysed. Chapter 3.4 explores the relationship between the oscillation behaviour and the vortex core trajectory using micromagnetic simulations.

### 3.1. Cases of Single-Domain and Vortex State in the Reference Layer

In the case where the reference layer is in a single-domain state, the hysteresis loop exhibits features correspond to the vortex core nucleation, displacement, and annihilation in the free layer [18], as shown in Fig. 1d. The field-cooling procedure aligns the RL magnetization along the –x direction, the application of a positive magnetic field $B_x$ aligns the FL magnetization antiparallel to the

RL (state 1), resulting in a high-resistance state. As the field is increased in the negative direction, vortex core nucleation occurs in the FL, leading to a sharp drop in resistance (state 2). Further increase in the negative field causes a quasi-linear displacement of the vortex core, eventually resulting in its annihilation at the disk edge. This transition is marked by a sudden resistance jump, indicating that the FL has transitioned into a single-domain state (state 3), which corresponds to a low-resistance state due to parallel alignment of magnetization in both layers. In Fig.1d, a pronounced offset of the loop can be observed which is due to the significant stray field produced by reference layer when in the single-domain state.

In the case of the vortex state in the RL, the hysteresis loop changes qualitatively (Fig. 1e) compared to the single domain state, specifically in terms of a strong dependence on the current applied to the MTJ. When applying a strong DC current, the resultant Oersted field will determine the vortex chirality in the FL during nucleation. Reversing the direction of the DC current changes the chirality of the Oersted field and, consequently, the vortex chirality. The chirality of the pinned vortex in the RL is set randomly during the annealing process. A configuration where the vortices in the RL and FL have the same chirality corresponds to the low-resistance state (2'), as the magnetizations in the adjacent layers are parallel. Conversely, a configuration with opposite chiralities corresponds to the high-resistance state (2), due to the antiparallel alignment of the magnetizations. The minimum current required to define the vortex chirality in the FL of a MTJ with a 1000 nm diameter is 8 mA, which generates an Oersted field of 3 mT at the periphery (S.2).

Saturation of the FL in a positive or negative external field results in an intermediate resistance value (1, 3), unlike the case where the RL is in a single-domain state, which produces distinct low- and high-resistance states. Additionally, there shift along the x-axis due to the stray field from the reference layer is significantly reduced due to the magnetic vortex having minimal stray field (apart from the vortex core).

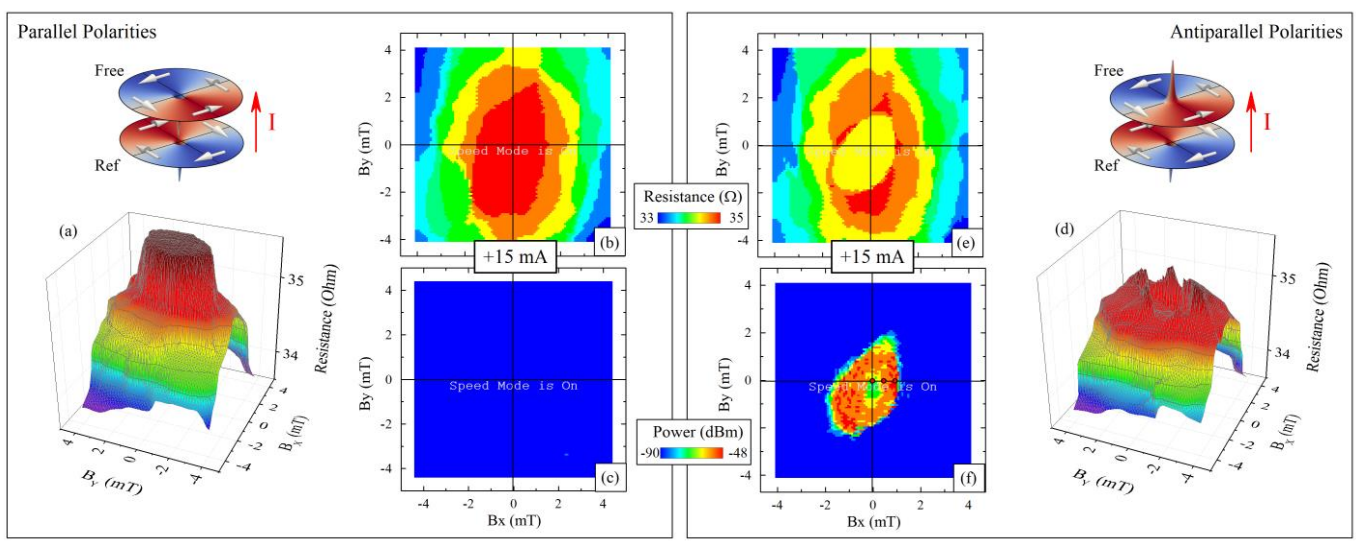

*Fig. 2.* Dependence of resistance and oscillation power on the vortex core position in the free layer, controlled by a pair of external fields Bx and By. In the case of parallel polarities of the vortices in the

*reference and free layers, core-to-core attraction results in a resistance plateau (a, b) and suppression of oscillations (c). In contrast, for antiparallel polarities, repulsion between the cores enables oscillations (f) and leads to a decrease in resistance in the oscillation region (d, e).*

### 3.2. Auto-Oscillation of a Vortex Core

The static and dynamic properties of the system were studied as a function of external magnetic fields in the presence of a +15 mA DC current. The application of Bx and By fields enables manipulation of the vortex core position within the plane of the FL disk. The field values were gradually increased from zero in a spiral pattern. For each field pair, three types of measurements were recorded: (1) average resistance using a source-measure unit (SMU); (2) time-dependent resistance using an oscilloscope; and (3) frequency spectrum using a spectrum analyzer. In the low-resistance state, where the vortices in the RL and FL have the same chirality, no auto-oscillations were observed. The results for the case of opposite chiralities and various combinations of core polarities are presented below.

For cases of parallel and antiparallel combinations of polarities, the resistance distributions as a function of external fields are shown in Fig.2 in both 3D (a,d) and 2D (b,e) representations. The distribution for both combinations shows some similarities, with a reduction in for larger fields applied in either the x or y direction. However, close to zero applied magnetic field there is a pronounced difference between the resistance profile for the two polarities, with a plateau (a) and a valley (d) being visible. This difference in behaviour can be understood in terms of the dipolar core to core coupling between the pinned vortex in the RL and the free vortex in the FL. In the case of similarly oriented core polarities there will be a strong pinning effect of the FL vortex core, where the vortex core is trapped by the presence of the magnetic field emanating from the RL, resulting in a plateau of similar resistance values as the vortex is not displaced by the external field. In the case of anti-parallel oriented vortex core polarities, the core in the FL will be repelled by the stray field from the RL. In fact the resultant valley in resistance can be understood by analysing the devices in the frequency domain (Fig.2c,f) where the peak power of the high frequency signal produced by the device is plotted as a function of the in-plane magnetic field. For the antiparallel core polarity alignment, the device can be seen to emit a non-zero rf power in the same region as the resistance.

The origin of the valley is related to the fact that the resistance measurements are time-averaged over many cycles. Averaging a quasi-circular orbit over multiple cycles results in a resistance level that is relatively lower than the static value. If the current passed through the system is insufficient to excite oscillations (e.g., 7 mA), then instead of a valley in the resistance distribution, a smooth decrease in resistance from the maximum value is observed as the field increases (see S.3). Increasing the current passed through MTJ leads to an increase in the oscillation area on the Bx-By diagram. The corresponding experimental results for currents from 10 to 16 mA are given in S.4.

To summarise, it was experimentally found that current propagation through the MTJ in the high-resistance state, characterized by two vortices with opposite chirality, can excite steady-state oscillations of the vortex core in the FL. A systematic study showed that the ability to sustain such oscillations is determined by the combination of vortex core polarities in the RL and FL. The interaction between the cores is attractive when their polarities are parallel and repulsive when they are antiparallel. Micromagnetic simulation allows to estimate the maximum field value above the RL core at 0.1 T (Fig.4a). Attraction prevents oscillations by pinning the vortex core in the FL directly above the vortex core in the RL. Conversely, repulsion promotes oscillation by eliminating the global energy minimum at the central position. The core polarity (+P/–P) is defined probabilistically during the nucleation process and is independent of vortex chirality. Under these conditions, auto-oscillations were detected in devices with diameters of 1000, 900 and 800 nm (S.5).

### 3.3. Features of Oscillation

When the core polarities are antiparallel and steady auto-oscillations in the free later vortex are induced, the nature of the oscillations are non-trivial. Visualization with a spectrum analyser and oscilloscope shows that the oscillation changes qualitatively with the external field. At the periphery of the oscillation region (Fig.2f at 1 mT), the time-domain signal appears as a sine wave with periodic disturbances at the same frequency and a fixed phase shift. As a result, the spectrum displays a main peak around 60 MHz and an additional peak at the second harmonic, around 120 MHz (Fig. 3a).

Reducing the field to 0.6 mT transforms the disturbances into additional oscillations (Fig. 3b). In the frequency domain, this is reflected as two peaks of equal height. A field of 0.3 mT corresponds to the centre of the oscillation region. Here, the increased contribution of the secondary oscillation results in a time-domain signal that appears as a pure second-harmonic oscillation, with a corresponding peak in the spectrum (Fig. 3c). Since resistance oscillations are associated with the movement of the vortex core in the FL, the complex signal is related to the deformation of the core's trajectory. These aspects will be discussed in detail in the next chapter.

To illustrate the overall impact of the external field on oscillation power and frequency, the dependencies of the first and second harmonic peaks are shown in Fig. 3d,e. For the first harmonic, the power reaches a minimum at the center of the oscillation region ($B_x$ = –0.3 mT) and increases toward the periphery, reaching a reflected power of approximately 3 nW. In contrast, the second harmonic exhibits the opposite trend. The oscillation frequency displays a complex profile, varying within a 10 MHz range, with minima observed both at the center and at the periphery of the oscillation region.

The influence of the applied current magnitude on the power and frequency of oscillations was additionally investigated (Fig. 3f,g) at a fixed magnetic field of 1 mT. Oscillations begin once the threshold current of 7 mA is exceeded. The oscillation power increases nonlinearly up to 15 mA for both peaks. The main frequency increases linearly by approximately 15 MHz as the current increases from 7 to 17 mA.

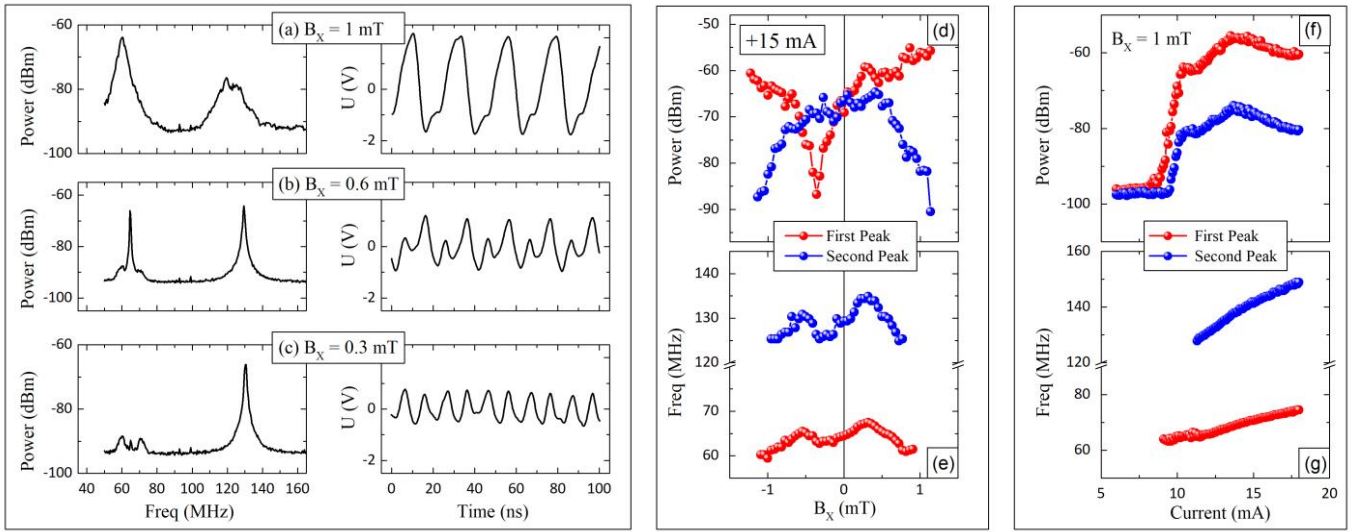

*Fig. 3.* (a-c) Examples of the oscillation spectrum and corresponding time-domain signals for different values of Bx. Dependence of oscillation power (d) and frequency (e), characterized by two peaks in the spectrum, on the external magnetic field applied along the x-axis. (f, g) Dependence of oscillation power and frequency on the current at a fixed external field.

### 3.4. Simulation of the vortex core trajectory

In order to better understand the complex nature of the dynamics of the auto-oscillations in the vortex free layer micromagnetic simulations were performed with the MuMax3 code [22] (Fig.4a), the file with the corresponding script can be found in the Supplementary Materials. Due to its large thickness, the core of the fixed vortex in the reference layer generates a magnetic field of approximately 0.1 T at the centre, which is sufficient to pin the core in the FL (Fig.4a). As a result, in the two-layer system, oscillations are observed only for one polarity. In contrast, in the three-layer system, where the RL is five times thinner, oscillations are observed for both polarities [arXiv]. Additionally, the simulations do not impose any polarity restrictions on the oscillations. The mechanism of auto oscillations will be explained in the next chapter, but for now the behaviour of the system resistance will be discussed.

Experimentally, the position of the vortex core is not measured directly. Instead, the resistance of the structure, arising from the tunnel magnetoresistance (TMR) effect [23], is measured. This resistance depends on the relative magnetization distribution in the reference layer (RL) and the free layer (FL), and it can be calculated from simulation results, as described in Section S.6.

In the absence of any symmetry breaking, the spin-polarized current drives the vortex core into a circular orbit cantered on the pillar (i.e., the red trajectory in Fig. 4a). In this case, the vortex core in the FL follows a path that is equidistant from the centre at all times, meaning that the system remains in a constant resistance state throughout the orbit. As a result, there is no measurable change in resistance, since all positions along the trajectory correspond to the same resistance value (Fig. 1b). Therefore, even if oscillations are present, they cannot be experimentally detected when the system and its dynamics are fully symmetric about the centre.

The presence of a constant external field shifts the orbit of the gyrotropic motion of the vortex core and states 1 and 3 cease to be symmetrical, the blue trajectory in Fig. 4. In this case, the dependence of the resistance reflects the presence of oscillations (Fig.4c). An increase in the external field leads to a further shift in the trajectory and, accordingly, to an increase in the amplitude of the resistance oscillations (Fig.4d). Whilst the symmetry breaking can explain why the peak at the main frequency is reduced to zero close when the two vortices are closely aligned (i.e. close to zero in-plane magnetic field) it does not explain the emergence of the second harmonic peak.

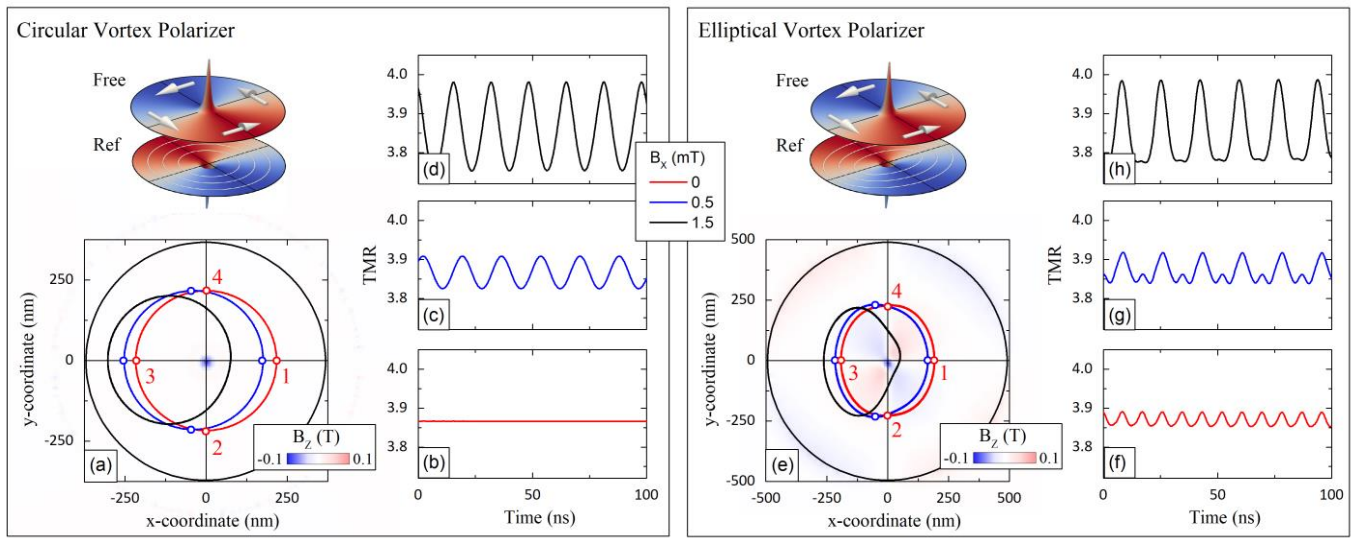

*Fig. 4.* Simulation of auto-oscillation driven by vortex-polarized spin current. In the case of a symmetrical vortex polarizer (a), the trajectory of the vortex core in the free layer is circular for all values of the external field, with a corresponding sinusoidal variation in tunnel magnetoresistance (b–d). Elliptical deformation of the vortex polarizer leads to an elliptical core trajectory in the free layer (e) and more complex resistance responses (f–h). The colormap in (a) and (e) represents the distribution of the magnetostatic field generated by the reference layer in the free layer.

In order to explain the emergence of the second harmonic peak it is necessary to introduce uniaxial deformation to the RL. The presence of uniaxial deformation causes both a change in the distribution of the magnetostatic field (colourmap in Fig.4e) and in the distribution of the spin current polarization. This uniaxial deformation can be introduced to the device assuming that the antiferromagnet pinning the RL has some residual anisotropy [arXiv], which is transmitted to the RL through exchange bias, however, further investigation is necessary to ascertain the exact nature of this anisotropy.

When the simulations are performed with a deformed RL, in the absence of an external field, the trajectory of the core becomes elliptical, the red line Fig.4e. In this case, state 1 and state 2 correspond to different resistance values. Another important consequence of ellipticity is that during one period, the state with low (1 and 3) and high (2 and 4) resistance is repeated twice. Due to the symmetry of these states, the change in resistance occurs at a frequency double (Fig.4f) that of the oscillations, which is observed experimentally.

The application of an external field leads to a shift of the elliptical trajectory relative to the centre, blue line. Previously symmetric states 1 and 3 cease to be such and two types of oscillations are designated, associated with the shift of the trajectory and with its ellipticity. Further increase of the field leads to significant deformation of the trajectory associated with repulsion between the cores, black line in Fig.4e. In this case, the dependence of the resistance has the features of a sinusoid with deformations (Fig.4h).

4. Qualitative Description of Auto-Oscillation Mechanisms

The phenomenon of vortex core auto-oscillation was previously observed in spin valves with an unfixed reference layer [24] [25] [26] [27] [28] [29] [30]. We also investigated auto-oscillations and provided an explanation of the mechanism in a three-layer MTJ with a so-called synthetic antiferromagnetic structure, where the reference layer is antiferromagnetically coupled to the pinned layer [arXiv]. In the present work, a two-layer MTJ was used, simplified by eliminating the pinning layer and increasing the thickness of the RL to 12 nm. This modification makes the formation of a single vortex beneath the FL more natural and pronounced, in contrast to the pair of vortices with opposite chirality discussed in the three-layer MTJ.

In the case of a vortex in the RL, the auto-oscillation mechanism can be qualitatively explained as follows. As current passes through the MTJ, it becomes spin-polarized, with the polarization distribution mirroring the magnetization distribution in the RL. The interaction between this vortex-polarized current and the magnetic vortex in the FL can be described by the Slonczewski spin-transfer torque [31] [32]. The reactive part is damping-like component of the torque, $\tau\_DL \propto [m \times [m\_p \times m]]$, where m_p is the spin current polarization and m is the magnetization of the FL [33] [34]. This interaction can be interpreted as the creation of an effective field, $B\_DL \propto [m\_p \times m]$, acting on the vortex. Since both the magnetization and polarization vectors lie in the plane of the disk, the effective field is directed perpendicular to the plane, as illustrated in the B_DL distribution, Fig.5a. Localized regions of this effective field can attract the vortex core. Displacement of the vortex core leads to a redistribution of the field, and a corresponding shift in regions of maximum B_DL, resulting in a sustained motion of the vortex under the influence of direct current (Fig.5). The effective field distribution was generated using the custom fields function in MuMax3; the script is provided in the Supplementary Material.

Vortex auto-oscillations can also be observed when the RL magnetization is oriented perpendicular to the plane [14] [35]. In this case, the spin current polarized along the z-axis induces an in-plane field, with a distribution shown in Fig.5b. It is seen that the field is directed orthogonally to the magnetization at each point of the vortex, which should not initiate its motion. However, averaging the field over the surface shows that it can be characterized by a resulting direction. The action of this resulting field leads to a shift in the vortex core and a subsequent change in the direction of the resulting field, which leads to stable oscillations (Fig.5b). During the motion of the core, the magnitude of the average field does not change, but its direction does.

By comparing both cases, it can be concluded the following. In the case of vortex-polarized current, oscillations arise due to the action of the perpendicular component of the effective field on the vortex core. In contrast, with perpendicularly polarized current, oscillations result from the in-plane resulting field acting on the planar magnetization of the vortex.

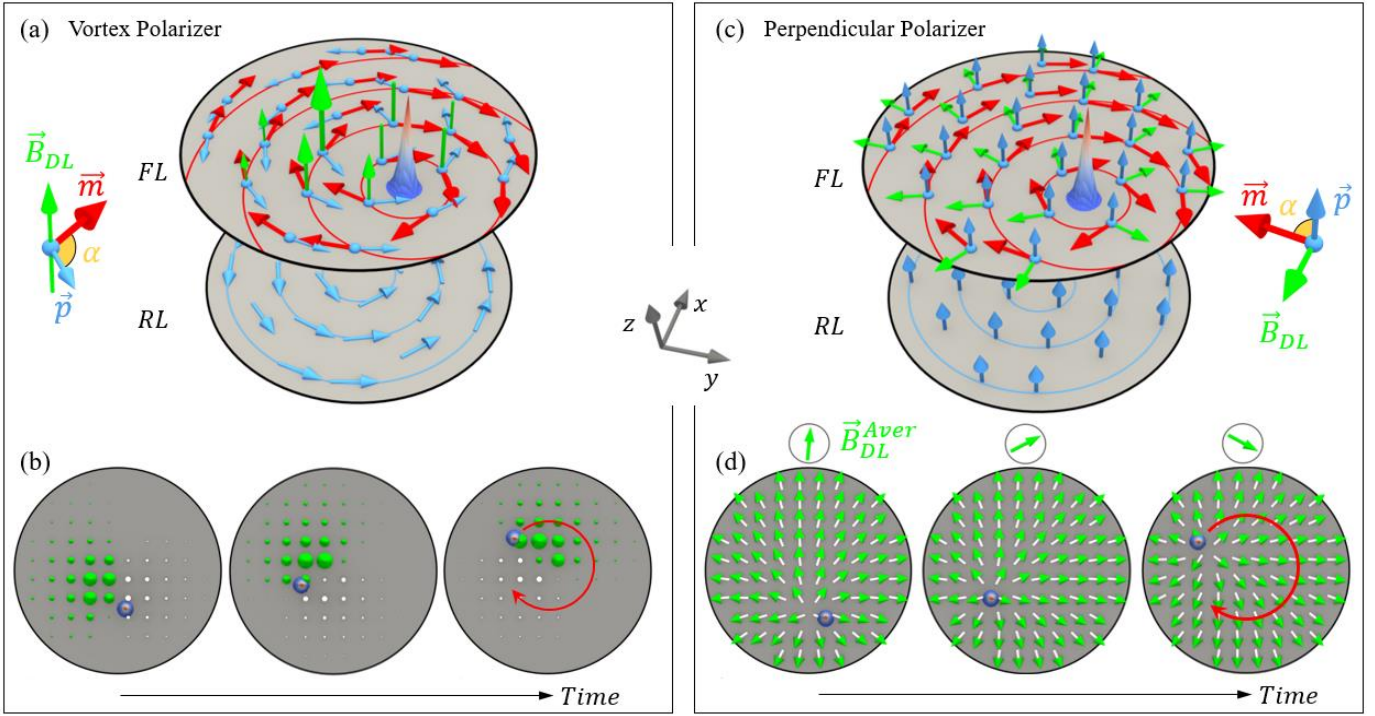

***Fig. 5.*** *Schematic illustration of the vortex auto-oscillation mechanism, showing the induction of an effective field BSTT as a result of the Slonczewski torque. The diagram compares the cases of vortex (a) and single-domain (c) states in the reference layer, which act as spin polarizers. The corresponding time evolution of the vortex core position and the associated distribution of the effective field are shown at the bottom (b, d).*

## 5. Conclusion

In this work, we experimentally demonstrate a magnetic vortex which is pinned directly via the coupling to an antiferromagnet and show how the dipolar coupling between the pinned vortex core and vortex in the free layer strongly modify the response of the free vortex. Steady state auto-oscillations are observed in the free layer vortex when the polarities and chiralities of the pinned and free vortices are anti-parallel. The effect is demonstrated in MTJs with diameters ranging from 1000 to 800 nm, exhibiting oscillation frequencies between 60 and 110 MHz.

A qualitative explanation of the oscillation mechanism is provided with the help of micromagnetic simulations. When a spin current with vortex-like polarization is injected, it generates local perpendicular effective fields due to spin-transfer torque, which drive vortex core motion. Observations in the time and frequency domains reveal how the first and second harmonic components vary with the in-plane field. This behavior can be explained by assuming a uniaxial deformation of the pinned vortex in the reference layer.

## 6. Materials and Methods

The structures under study were obtained based on a film of the following composition from bottom to top: [Ta(5)/CuN(25)]x6/Ta(5)/Ru(5)/IrMn(10)/CoFe(2)/CoFeSiB(16)/CoFeB(2.5)/MgO/CoFeB(2)/Ta(0.21)/NiFe(7)/TiWN(15)/AlSiCu(200)/TiWN(15), thicknesses in nm. The film was deposited on the surface of the 200 mm thermally oxidized wafer Si/SiO2(200 nm) using magnetron sputtering Singulus TIMARIS Multi-Target-Module. The layers below the first IrMn layer are further used to form the bottom electrical contact and are optimized to minimize roughness. The layers above the NiFe layer are needed to protect the magnetic stack from numerous further etching procedures. Then, nano-pillars with a diameter from 100 to 1000 nm were formed using electron beam lithography and ion beam milling. The nano-pillar was insulated with SiO2, after which the bottom and top electrical contacts were formed, as shown in Fig.1a. The thickness of the MgO layer was selected to obtain RxA=9 Ωμm2. In this case, the tunnel magnetoresistance is 120% at a reading current of 0.1 mA. When the reading current increases to 10 mA, the value decreases to 45%. After fabrication, the structures were annealed at 330 °C for 2 hours in the absence of a field or in the presence of Bx = 1 T. This procedure allows fixing the magnetization in the reference layer in a vortex or single-domain state due to the effect of exchange bias at the boundary with IrMn.

The magnetic properties of the stack were studied using a vibrating sample magnetometer (VSM) from MicroSense. The magnetotransport properties were investigated using a probe station equipped with an electromagnet generating an in-plane magnetic field $B_y$ with an amplitude of up to 0.1 T and a two-point high-frequency probe. To generate a magnetic field along the x-axis, a field line fabricated over the MTJ column was used (Fig. 1a), enabling the creation of magnetic fields up to 40 mT with an efficiency of 0.2 mT/mA. The hysteresis loop was recorded using a Keysight B2901B source measurement unit (SMU), which measured the resistance of the structure during magnetic field sweeps along the x- or y-axis. Auto-oscillations were detected using an Agilent E446A spectrum analyzer and an Agilent DSO-X 92004A oscilloscope, both connected through a bias-tee (Fig. 1a).

## 7. Acknowledgments

This work has received funding from the European Union's Horizon 2020 research and innovation programme under grant agreement No 101017098 (project RadioSpin), No 899559 (project SpinAge) and No 101070287 (project Swan-on-chip) and No 101070290 (project NIMFEIA).

## Supplementary Materials

### S.1. Magnetic Properties Of The Film

The film with the composition IrMn(10)/CoFe(2)/CoFeSiB(16)/CoFeB(2.5)/MgO/CoFeB(2)/Ta(0.21)/NiFe(7) was annealed for two hours at a temperature of 330 °C in an in-plane magnetic field of 1 T to fix the single-domain state in a specific direction in both the reference and reference layers (Fig. S1a). Figure S1b shows a representative hysteresis loop for the film, measured using a vibrating sample magnetometer (VSM) at room temperature. Key stages of magnetization reversal are marked on the loop. In state (1), the magnetization in all layers is aligned with the external negative magnetic field. During the transition from state (1) to state (2), the magnetization in the pinned layer rotates due to the effect of exchange bias at the interface with the antiferromagnetic (AFM) layer. An increase in the applied field in the positive direction causes switching of the free layer (state 3).

Based on the hysteresis loop, the saturation magnetization ($M_s$) and indirect exchange energy ($J_{ex}$) for each layer were calculated [53], as shown in Fig. S1b. The saturation magnetization is 0.67 MA/m for the free layer, 0.8 MA/m for the reference layer.

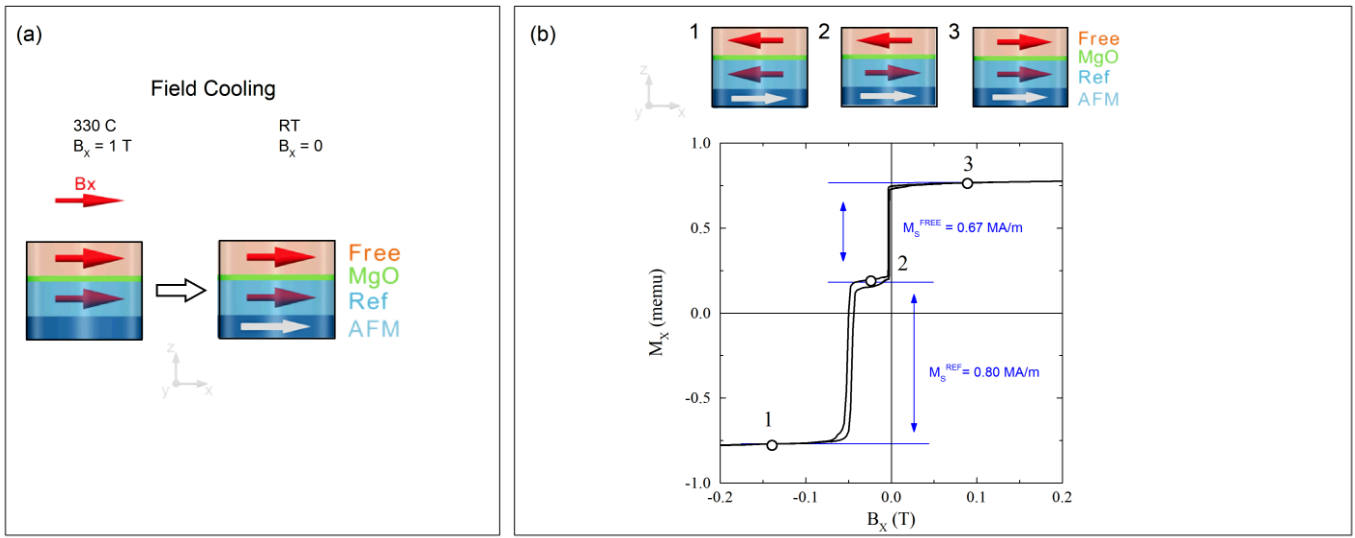

**Fig.S1.** *(a) Field cooling procedure. The sample was annealed at 330°C for three hours in a 1 T field along the x-axis and then cooled in the same field. (b) The magnetic hysteresis loop was obtained from a continuous film 6 × 8 mm using a vibrating sample magnetometer. The stages of magnetization reversal are marked with numbers and corresponding diagrams.*

### S.2. Oersted Field Estimation

The propagation of a current generates an Oersted field. For a quantitative estimation of the relationship between the current and the resulting magnetic field, an analytical model was used.

Within the conductor, the field distribution follows a circular symmetry, as described by relation $B = \frac{\mu_0 r}{2\pi R^2}$, which is obtained from the Biot–Savart law. The amplitude of the magnetic field remains constant at a fixed radius, and its orientation at each point is tangential to the corresponding circle (Fig. S2a).

The Oersted field is zero at the center of the MTJ column and increases linearly toward the periphery (Fig. S2b). For a column with a diameter of 1000 nm, the maximum field reaches approximately 3 mT for a 8 mA current, which is the minimum for specifying the chirality of the vortex, and 6 mT for a 15 mA current (Fig. S2b).

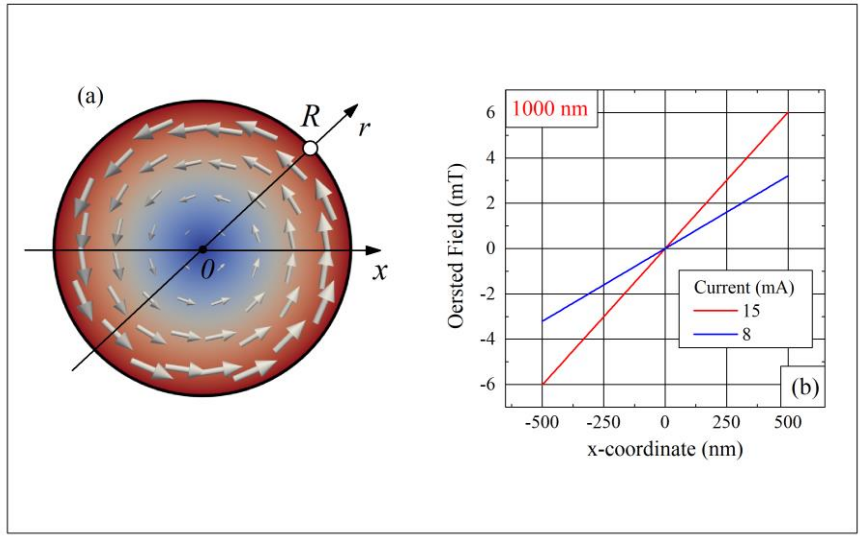

***Fig.S2.*** *(a) Schematic diagram of the Oersted field distribution in the disk when a current is passed. (b) Calculated dependence of the Oersted field in a disk with a diameter of 1000 nm on the x-coordinate for two different current values.*

### S.3. ResistanceProfile_NoOscillation

The onset of self-oscillations is accompanied by a decrease in the time-averaged resistance of the structure. To study the resistance profile as a function of the vortex core position in the free layer, measurements were performed at a constant current of 7 mA, which does not induce self-oscillations (Fig.S.3). It is evident that for parallel vortex polarities, a plateau appears: the vortex does not shift due to the strong interaction between the cores in the free and reference layers. In contrast, with antiparallel polarities, a smooth decrease in resistance is observed, reaching its maximum at the point corresponding to zero external field.

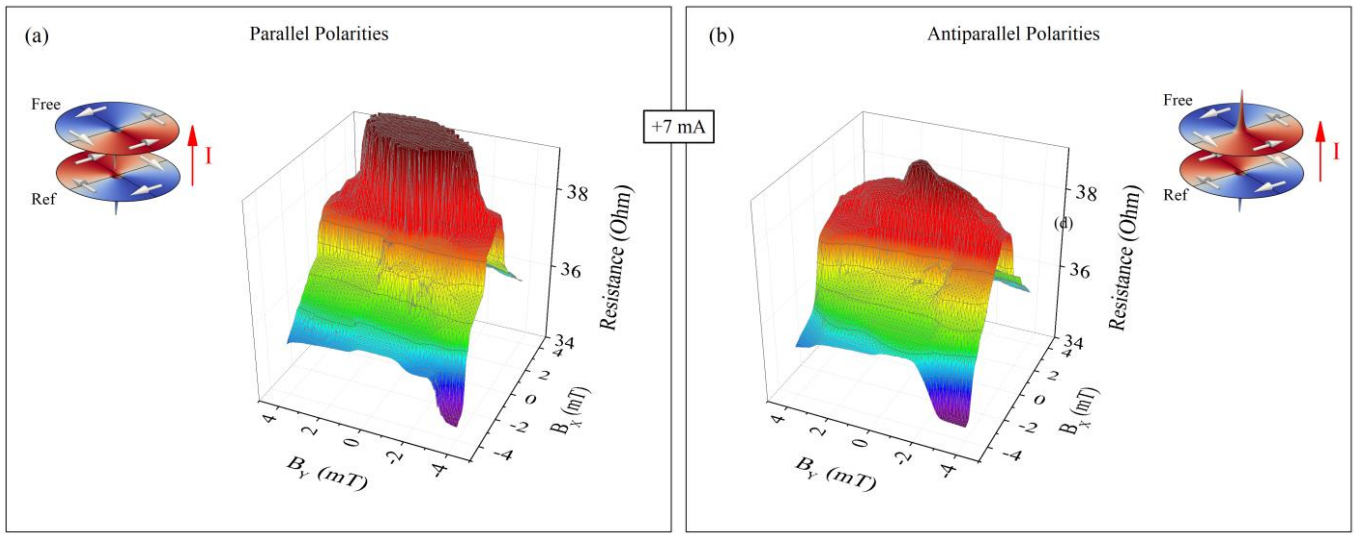

**Fig.S3.** *Distribution of resistance of a 1000 nm diameter MTJ in cases of parallel (a) and antiparallel (b) polarity of cores. The distributions were obtained in the presence of a current of 7 mA, which does not cause oscillations.*

S.4. Oscillations for Different Current

The behavior of the resistance distribution and the regions of auto-oscillations were studied as functions of the DC current (Fig.S4). In a device with a 1000 nm diameter, increasing the current from 10 to 16 mA leads to a significant expansion of the auto-oscillation region in the external field coordinates along both x and y. The area of the resistance dip increases accordingly.

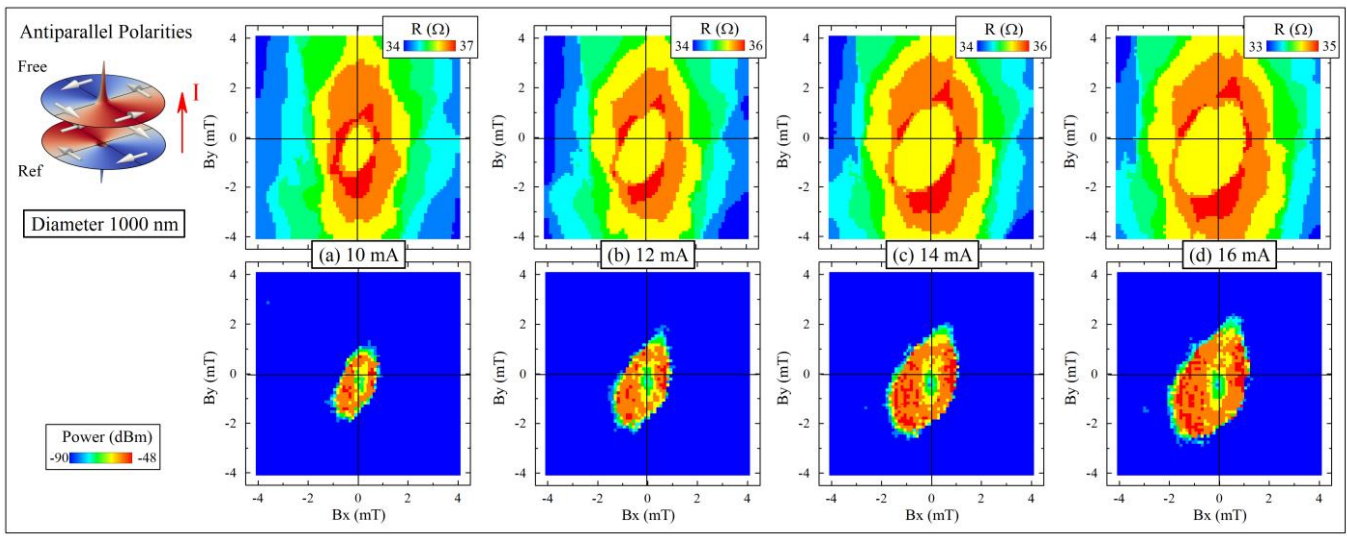

**Fig.S4.** *Experimentally obtained distribution of resistance and oscillation power for 1000 nm MTJ as a function of the $B_x$ and $B_y$ fields for different DC currents.*

S.5. Oscillations for Different MTJ diameters

The main text of the article is dedicated to describing the auto-oscillation and reconfiguration processes observed in a 1000 nm diameter MTJ. Nevertheless, all conclusions remain valid for devices with smaller diameters, as confirmed experimentally. Fig.S5. shows examples of resistance distribution and auto-oscillation regions in devices with diameters of 900 and 800 nm.

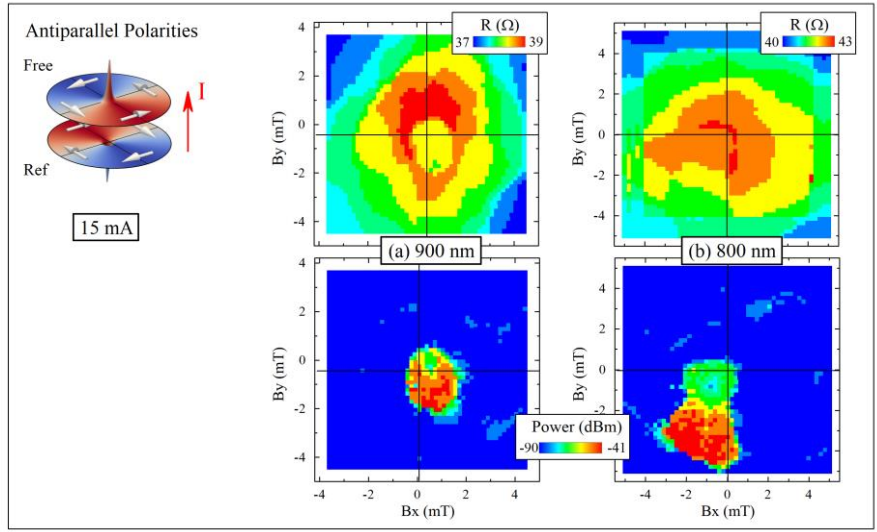

***Fig.S5.*** *Experimentally obtained distribution of resistance and oscillation power for 900 nm (a) and 800 nm (b) MTJ as a function of the $B_x$ and $B_y$ fields.*

S.6. Method for estimating tunnel magnetoresistance

During the two-dimensional scanning of the external field along the x- and y-axes, the vortex in the free layer shifts relative to the fixed magnetic structure of the reference layer. The resulting resistance distribution allows us to make indirect assumptions about the fixed magnetic structure. To confirm the assumptions made, experimental results are compared with the simulation results.

Tunnel magnetoresistance is directly proportional to the cosine [54] of the angle between the magnetic moments in the free and reference layers, Fig. 1. Using the results of micromagnetic modeling, the resistance was calculated for each pair of cells in the free and reference layers. Then these values were summed up and divided by the number of cells. As a result, the resistance value was obtained for a specific pair of magnetic states in the free and reference layers.

To construct the resistance distribution as a function of external fields, magnetic states of the vortex in the free layer were simulated at field values Bx and By ranging from ±5 mT, resulting in 400 distinct states. For each state, the resistance was calculated based on the magnetic configuration fixed in the reference layer. This produced resistance distributions corresponding to various recorded configurations in the reference layer (Fig.S6a).

When a symmetric vortex is fixed, the resistance distribution is also strictly symmetrical (Case 1). Deformation of the vortex along a single axis results in an elliptical resistance distribution (Case 2).

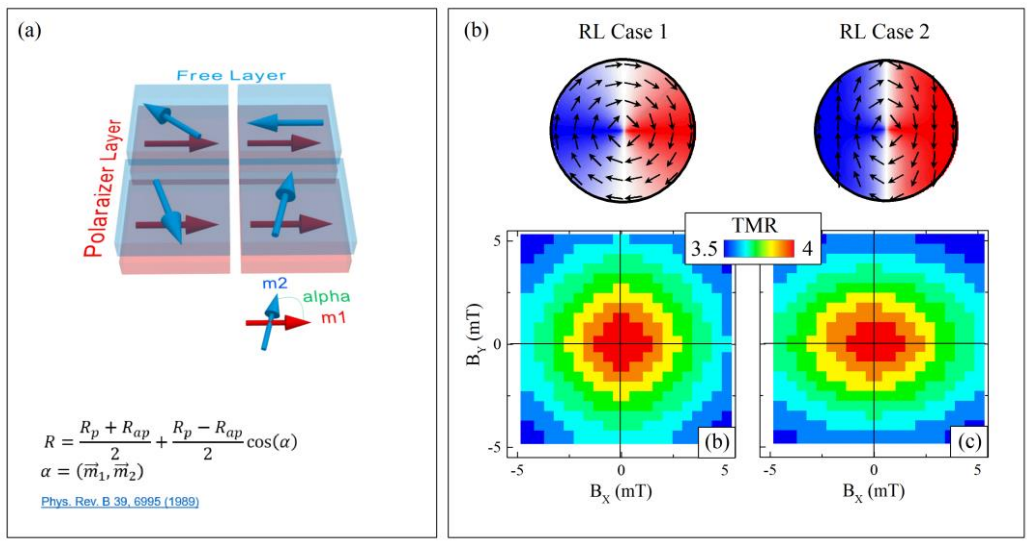

***Fig.S6.*** *(a) Schematic description of the method used to calculate the magnitude of resistance change due to the tunnel magnetoresistance effect in a structure divided into cells. Examples of resistance distribution in a structure where the sureference layer has a symmetric magnetic vortex (b) or a deformed vortex (c), and the symmetric magnetic vortex with opposite chirality in the free layer is shifted depending on the applied fields Bx and By.*